\documentclass[12pt]{iopart}

\usepackage[colorlinks=true, pdfstartview=FitV, linkcolor=red, citecolor=blue, urlcolor=blue]{hyperref}
\usepackage{bm}
\usepackage{cite}
\usepackage{graphicx}
\begin{document}

\title[Prompt, pre-equilibrium, and thermal photons in relativistic nuclear collisions]{Prompt, pre-equilibrium, and thermal photons in relativistic nuclear collisions}

\author{Akihiko Monnai$^{1, 2}$}

\address{$^1$ Department of Mathematical and Physical Sciences, Faculty of Science, 
Japan Women's University, Tokyo 112-8681, Japan}
\address{$^2$ KEK Theory Center, Institute of Particle and Nuclear Studies, 
High Energy Accelerator Research Organization (KEK),
1-1, Ooho, Tsukuba, Ibaraki 305-0801, Japan}
\ead{monnaia@fc.jwu.ac.jp}
\vspace{10pt}
\begin{indented}
\item[]April 2020
\end{indented}

\begin{abstract}
The direct photon emission model in relativistic nuclear colliders has been improved in recent years for reducing the discrepancy between theoretical estimations and experimental data and for understanding the properties of the QCD matter. In this study, the contribution of pre-equilibrium photons are investigated in addition to those of prompt and thermal photons in the framework of a relativistic hydrodynamic model. The numerical simulations at an LHC energy suggest that the pre-equilibrium photons may be relevant at intermediate transverse momentum near the saturation momentum scale, increasing particle spectra and reducing elliptic flow of direct photons.
\end{abstract}

%
%
%
%
%

\section{Introduction}
\label{sec1}

The QCD matter created in relativistic nuclear colliders such as the BNL Relativistic Heavy Ion Collider (RHIC) and the CERN Large Hadron Collider (LHC) has been revealed to be a strongly-coupled system that follows hydrodynamic description \cite{Wang:2016opj}. The quark-gluon plasma (QGP) \cite{Yagi:2005yb} and the subsequently-produced hadronic matter are considered to participate in thermal equilibrium before kinetic freeze-out. Electromagnetic probes such as photons and dileptons, on the other hand, do not interact with the QCD medium once they are produced because they do not have color charges. It is believed that they are useful in quantifying macroscopic space-time evolution as well as microscopic properties of the medium. The transverse momentum spectra of photons have been used to estimate the effective medium temperature from the inverse slope parameter; $T \sim$ 220-240~MeV at RHIC \cite{Adare:2008ab,Adare:2014fwh} and $T \sim$~300 MeV at LHC \cite{Wilde:2012wc,Adam:2015lda}. This, using the knowledge of the crossover temperature $T_c \sim$~160-170 MeV estimated in (2+1)-flavor lattice QCD \cite{Borsanyi:2013bia,Bazavov:2014pvz}, is considered as one of the evidences for the creation of the QGP in relativistic heavy-ion collisions.

Inclusive photons consist of direct photons from primary sources and decay photons from the subsequent hadronic decay processes. Conventionally, direct photons are estimated to be the sum of prompt photons, which are produced in the hard process at the time of the collision, and thermal photons, which are emitted from the soft, thermalized sector of the medium. 
This, however, may be an oversimplified picture because the relativistic nuclear collisions go through several stages, from initial color glass condensate to glasma, to hydrodynamic QCD matter, and to hadronic gas. Thus one has to consider the photon contributions from the pre-equilibrium stage \cite{Ornik:1995yk,Wang:1996yf,Blaizot:2011xf,Chiu:2012ij,McLerran:2014hza,Berges:2017eom,Oliva:2017pri,Khachatryan:2018ori,GarciaMontero:2019yoj} as well as from the post-hydrodynamic stage \cite{Baeuchle:2009ep,Schafer:2019edr,Schafer:2020vvw} for a comprehensive picture. 

The phenomenological description of direct photons, unlike that of hadrons, has not been well-established. Direct photons are known to have the photon puzzle where the observed azimuthal momentum anisotropy, quantified by elliptic and triangular flow, is larger than the estimations of the relativistic hydrodynamic model \cite{Adare:2011zr, Adare:2015lcd, Acharya:2018bdy}. The situation has been improved in recent years owing to the efforts made in refining the photon estimation model \cite{Turbide:2005bz,Chatterjee:2005de,Chatterjee:2008tp,Chatterjee:2011dw,vanHees:2011vb,Dion:2011pp,Tuchin:2012mf,Basar:2012bp,Liu:2012ax,Muller:2013ila,Linnyk:2013wma,Monnai:2014kqa,McLerran:2014hza,vanHees:2014ida,Vujanovic:2014xva,Monnai:2014taa,Gale:2014dfa,Monnai:2015qha,McLerran:2015mda,Paquet:2015lta,Linnyk:2015rco,Vovchenko:2016ijt,Koide:2016kpe,Iatrakis:2016ugz,Fujii:2017nbv,Ayala:2017vex}. Nevertheless, it is still considered to be apparent at RHIC and is shown as a systematic tendency at LHC albeit the statistics may need improvement for further confirmation for the latter case. Furthermore, the direct photon particle spectra tend to overshoot theoretical predictions. 
The existence of large direct photon triangular flow \cite{Adare:2015lcd} implies that the large momentum anisotropy is originated in the medium properties rather than in the external sources. 

In this paper, I aim for a proof-of-principle study and introduce the pre-equilibrium photon contribution in the framework of relativistic hydrodynamic modeling to eliminate the no-photon-emission stage of glasma in nuclear collisions for the first time. The pre-equilibrium photon emission is then compared with the conventional emissions of prompt and thermal photons for order estimate. An example for such proof-of-principle study can be found in Ref.~\cite{Denicol:2018wdp} where a single-shot hydrodynamic model is used for examination of baryon diffusion with bulk viscosity turned off. A full comparison to the experimental data is thus beyond the scope of the current study. More quantitative estimations involving data analyses and comparison of different pre-equilibrium models will be discussed elsewhere.

Self-similar scaling solutions are employed for the parton phase-space distributions in the pre-equilibrium stage as a model of photon emission \cite{Berges:2013eia,Berges:2014bba,Berges:2015ixa,Berges:2017eom,Tanji:2017suk} assuming that turbulent thermalization mechanism drives the early time dynamics in high-energy nuclear collisions at LHC. The process is integrated into the hydrodynamic model by scaling it into the typical time scale of local equilibration required by the hydrodynamic model. The inevitable presence of pre-equilibrium photons is na\"{i}vely expected to improve the total photon yield estimation while worsening the elliptic flow agreement in heavy-ion collisions. Thus it should be noted again that the main aim of the paper is not to completely resolve the photon puzzle but to establish a more complete picture of photon emission in nuclear collisions. It can also be relevant in proton-proton collisions because the expected production of glasma in high-luminosity events can modify the bottom-up prompt photon estimation from the experimental data that is often used as the baseline for thermal photon analyses in heavy-ion collisions \cite{Turbide:2003si}.

In Sec.~\ref{sec2}, the emissions of thermal, pre-equilibrium, and prompt photons are discussed analytically. They are used for the numerical estimation of direct photon spectra in Pb-Pb collisions at an LHC energy in Sec.~\ref{sec3}. The interplay of the photons emitted from the three different stages is discussed for transverse momentum spectra and differential elliptic flow. Sec.~\ref{sec4} is devoted for discussion and conclusions. The natural units $c = \hbar = k_B = 1$ and the mostly-minus Minkowski metric $g^{\mu \nu} = \mathrm{diag}(+,-,-,-)$ are used in the paper.

\section{Photon Emission Models}
\label{sec2}

The emission rates of thermal and pre-equilibrium photons as well as the spectra of prompt photons are discussed for the phenomenological estimation of direct photons in high-energy nuclear collisions. 

\subsection{Thermal photons}

The thermal photon emission rate is estimated as an interpolation of the hadronic and the QGP emission rates as a function of the temperature,
\begin{eqnarray}
E \frac{dR^{\gamma}_\mathrm{th}}{d^3p} &=& \frac{1}{2}\bigg(1- \tanh \frac{T-T_c}{\Delta T} \bigg) E \frac{dR^{\gamma}_\mathrm{had} }{d^3p} + \frac{1}{2}\bigg(1+ \tanh \frac{T-T_c}{\Delta T}\bigg) E \frac{dR^{\gamma}_\mathrm{QGP}}{d^3p} , \label{eq:th}\nonumber \\
\end{eqnarray}
where $T_c = 170$ MeV and $\Delta T = 0.1 T_c$. The hadronic photon emission rate includes the contributions of the processes in Ref.~\cite{Turbide:2003si,Heffernan:2014mla,Holt:2015cda}. The QGP photon emission rate is based on the perturbative QCD (pQCD) calculations~\cite{Arnold:2001ms}.

The relativistic hydrodynamic model is used for the estimation of background medium evolution. The energy in the emission rate is shifted by the Lorentz boost $E \to p^\mu u_\mu$ where $u^\mu$ is the flow.

\subsection{Pre-equilibrium photons}
\label{sec:preeq}

The pre-equilibrium stage in nuclear collisions is not well known. Here a phenomenological approach is conjectured similarly to Ref.~\cite{Berges:2017eom} but with integration to the hydrodynamic modeling in mind. It is assumed that the glasma phase is divided into three stages following the bottom-up scenario \cite{Baier:2000sb}: (a) the early stage $\tau_0 < \tau < \tau_1$ where hard partons dominate, (b) the intermediate stage $\tau_1 < \tau < \tau_2$ where the gluon density is less than unity, and (c) the late stage $\tau_2 < \tau < \tau_3$ where soft partons dominate. Here $\tau_0 = c_0 Q_s^{-1}$, $\tau_1 = c_1 Q_s^{-1} \alpha_s^{-3/2}$, $\tau_2 = c_2 Q_s^{-1} \alpha_s^{-5/2}$, and $\tau_3 = c_3 Q_s^{-1} \alpha_s^{-13/5}$,
where $Q_s$ is the saturation momentum scale, which originates in the color glass condensate picture \cite{McLerran:1993ni,McLerran:1993ka,Iancu:2003xm,Gelis:2010nm,Kovchegov:2012mbw}. The coefficients $c_{0,1,2,3}$ are introduced to linearly scale the equilibration process into the time scale of pre-equilibrium evolution implied by the relativistic hydrodynamic modeling of nuclear collisions, $\tau_\mathrm{ini} \leq \tau \leq \tau_\mathrm{hyd}$. Typically, $\tau_0 = \tau_\mathrm{ini} \sim \mathcal{O}(10^{-1})$ fm and $\tau_3 = \tau_\mathrm{hyd} \sim \mathcal{O}(10^{0})$ fm. 

Following the procedure in Ref.~\cite{Berges:2017eom, Blaizot:2014jna}, the pre-equilibrium photon emission rate is given as
\begin{eqnarray}
E \frac{dR^\gamma_a}{d^3p} &=& \frac{20}{9\pi^2} \alpha_\mathrm{EM} \alpha_s f_q(p) \log \bigg(1+\frac{2.919}{g^2}\bigg) \int \frac{d^3p'}{(2\pi)^3} \frac{1}{p'} [f_g(p') + f_q(p')], 
\end{eqnarray}
where the Coulomb logarithmic factor is chosen so that the expression is a non-equilibrium extension of the well known result for quark pair annihilation and quark-gluon Compton scattering \cite{Kapusta:1991qp,Baier:1991em}.

A prominent approach to the early pre-equilibrium stage (a) is the classical statistical method \cite{Berges:2013eia,Berges:2014bba,Berges:2015ixa}. The quark and gluon distribution functions are implied to be described by the universal self-similar scaling law as
\begin{eqnarray}
f_g &=& (Q_s \tau)^\alpha f_{g}^{s} ((Q_s \tau)^{\beta} p_T,(Q_s \tau)^{\gamma} p_z), \label{eq:preeq_dist1} \\
f_{g}^{s} (p_T,p_z) &=& A_g p_T^{-1} \exp(-p_z^2/\sigma_{z}^2), \label{eq:preeq_dist2}
\end{eqnarray}
for gluons and similarly with the subscript change $g \to q$ for quarks \cite{Tanji:2017suk} in the classical regime. $p_T$ and $p_z$ are the transverse and the longitudinal momenta perpendicular and parallel to the collision axis, respectively. The exponents are $\alpha = -2/3$, $\beta = 0$, and $\gamma = 1/3$ in the intermediate $p_T$ range below $Q_s$.
The distributions are cut off for $p_T > Q$ with a hyperbolic tangent function to imitate the situation where it falls off fast with a different scaling law. This is introduced as an envelop function $\{1-\tanh[(p_T-Q_s)/\Delta p_T]\}/2$ where $\Delta p_T = 0.1 Q_s$ on the phase space distribution.

The coefficients $A$ and $\sigma_{z}$ are non-universal. Here $\sigma_z$, which is related with the momentum anisotropy of the system and can be interpreted as the longitudinal ``temperature", is treated as a parameter common to quarks and gluons and the normalization $A_g$ and $A_q$ are constrained with the energy density estimation explained later as functions of $Q_s$. $p_T$ dependence of $\sigma_z$ is neglected for simplicity unless otherwise mentioned. The procedure is made partly because the parameters are dependent on the initial conditions, extrapolated to the strong-coupling regime, and scaled so that the equilibration process fit into the time window allowed by the hydrodynamic model. 

The initial energy density distribution to the pre-equilibrium stage is assumed to be given by the Glauber model. The overall normalization is determined so that the energy density distribution at the end of pre-equilibrium evolution serves as the initial condition for the hydrodynamic model which in turn describes the experimental data after hadronic decay. The energy density in the pre-equilibrium stage estimated as
\begin{eqnarray}
e &=& \frac{1}{(2\pi)^3} \int_0^\infty 2 dp_z \int_0^{\infty} 2 \pi p_T \sqrt{p_T^2 + p_z^2} (d_g f_g + d_q f_q) , \nonumber
\end{eqnarray}
where $d_g = 2_\mathrm{spin}\times(N_c^2-1)$ and $d_q = 2_\mathrm{spin}\times 2_\mathrm{q\bar{q}}\times N_c \times N_f$. Assuming the quark density at early times is suppressed by $\alpha_s$ compared with the gluon density, \textit{i.e.}, $A_q = \alpha_s A_g$, the normalization factors are determined. When the longitudinal pressure is small enough compared with the transverse one, the internal energy density has an asymptotic form
\begin{eqnarray}
e &\sim& \frac{(d_g A_g + d_q A_q) Q_s \sigma_z}{8 \pi^{3/2} \tau}, \label{eq:ene_asymptotic}
\end{eqnarray}
which implies that it is proportional to the inverse of the time. This is subject to the correction from the soft sector in the stage (c) as mentioned later. It is simply assumed that the expression is valid just before the hydrodynamic stage because the exact thermalization dynamics is not well known and the proper time dependence of the energy density changes at most to $\tau^{-4/3}$ only for a relatively short period of time. 

The gluon number density is  
\begin{eqnarray}
n_g &=& \frac{d_g}{(2\pi)^3} \int_0^\infty 2 dp_z \int_0^{\infty} 2 \pi p_T f_g \sim \frac{d_g A _g \sigma_z}{4 \pi^{3/2} \tau} ,
\end{eqnarray}
and similarly for the quark number density with the change of the subscript $g\to q$. The number densities only have a weak dependence on the choice of $\sigma_z$ since the energy density is fixed in Eq.~(\ref{eq:ene_asymptotic}). Also the result is in qualitative agreement with the bottom-up estimation of the gluons freed at early time $n_g = c (N_c^2-1) Q_s^2 / 4 \pi^2 N_c \alpha_s \tau$ where $c$ is a dimensionless constant of $\mathcal{O}(1)$ \cite{Mueller:1999pi,Baier:2000sb} for a reasonable choice of parameters. 

The expression is assumed to be still valid for quarks in the intermediate stage (b). The occupation number for gluons decreases and becomes smaller than unity. The Debye screening of mass starts to be controlled by soft gluon contributions and the soft gluon number behaves as $\tau^{-1/2}$. On the other hand, the total number of gluons are still dominantly contributed by hard gluons. It is simply assumed here that the emission rate of the stage (a) can be extrapolated. 

In the late equilibration stage (c), the system is expected to transit from a non-thermal one to a thermal one. While the details of the mechanism is yet to be fully unveiled, a successful description should smoothly connect the photon emission rates as well. Here it is simply given by the interpolation 
\begin{eqnarray}
E \frac{dR_c^\gamma}{d^3p} &=& \frac{\tau - \tau_2}{\tau_3 - \tau_2} E \frac{dR^\gamma_\mathrm{th} }{d^3p} + \frac{\tau_3 - \tau}{\tau_3 - \tau_2} E \frac{dR^\gamma_b}{d^3p} .
 \end{eqnarray}
This also imitates the dominance of the soft sector and the build-up of the thermal tail in the hard sector in the emission rate. At $\tau = \tau_3 = \tau_\mathrm{hyd}$, the thermal emission rate (\ref{eq:th}) is recovered.
 
\subsection{Prompt photons}
 
Prompt photons are conventionally assumed to be the same as the scaled direct photons in proton-proton collisions. This is supported by the pQCD calculation of the photon yields, though it is not clear how valid such calculations are in low momentum regions. 

The direct photon spectra of proton-proton collisions is scaled with the number of collisions $N_\mathrm{coll}$ and parametrized as
\begin{eqnarray}
E \frac{dN^\gamma_\mathrm{scaled}}{d^3p} &=& 6745 \frac{\sqrt{s}}{(p_T)^5} \frac{N_\mathrm{coll}}{\sigma_{pp}^\mathrm{in}},\label{eq:ppscaled}
\end{eqnarray}
where $\sigma_{pp}^\mathrm{in}$ is the inelastic nucleon-nucleon cross section in units of pb \cite{Turbide:2003si}. The direct photons are assumed to be primarily prompt photons for the moment. Possible effects of the non-prompt photon contributions in proton-proton collisions \cite{Monnai:2018eoh} are discussed in \ref{sec:A}. 
Since the emission is supposed to be instantaneous, the prompt photon emission is given in the form of a number distribution rather than a rate.

\section{Numerical analyses}
\label{sec3}
The direct photon emission in relativistic nuclear collisions is estimated. Pb-Pb collisions at $\sqrt{s_{NN}} = 2.76$ TeV are considered. The inelastic nucleon-nucleon cross section is set to $\sigma_{pp}^\mathrm{in} = 65$ mb. The system is assumed to be net baryon free at this energy, though extension to the finite density system is a straightforward task.

Oftentimes, the pre-equilibrium stage in nuclear collisions is discussed assuming it is a homogeneous plane in the transverse directions. Here, on the other hand, the initial transverse distribution of the energy density in the pre-equilibrium stage is given by the Monte-Carlo Glauber model \cite{Miller:2007ri}. The initial time is set to $\tau_\mathrm{ini} = 1/Q_s$, \textit{i.e.}, $c_0 = 1$. The average $Q_s$ is expected to be around 2-3 at the LHC energies, which is motivated by the estimation $Q_s^2 \sim A^{1/3} Q_0^2 (x_0/x)^\lambda$ where $Q_0 = 1$ GeV, $\lambda = 0.288$, $x_0 = 3\times 10^{-4}$ for $x \sim 5\times 10^{-4}$ \cite{GolecBiernat:1998js}.
In this demonstrative study, the initial condition is event-averaged with the fixed impact parameter $b=4.6$ fm. This roughly corresponds to the average impact parameter for the 0-20\% centrality events. 
The averaged number of collisions $N_\mathrm{coll} = 1256$ is used. 
A pure longitudinal and boost-invariant expansion is assumed during time evolution. The contribution from the volume elements where the local temperature at the time of hydrodynamization is larger than the kinetic freeze-out temperature $T_f$ \cite{Cooper:1974mv} is taken into account. $T_f = 140$ MeV is used here and in the hydrodynamic model mentioned later.

The value of $\sigma_z$ depends on the initial dynamics of the collision but is expected to be smaller than a typical thermodynamic temperature scale of the given energy because the longitudinal pressure is small in the glasma picture. Here it is conjectured to be smaller than 0.5~GeV to ensure the condition $P_T > P_Z$.

The initial condition for the hydrodynamic stage is constructed directly from the energy density distributions at the end of the pre-equilibrium stage as the normalization is already constrained with the experimental data. The hydrodynamization time is chosen as  $\tau_\mathrm{hyd} = 0.6$ fm/$c$. 
The equation of state is based on the lattice QCD estimations \cite{Bazavov:2014pvz} connected with the hadron resonance gas model results. See Ref.~\cite{Monnai:2019hkn} for details. The viscosity is chosen to be vanishing for the moment because there remains ambiguity in the treatment of viscous corrections to thermal photons and the overall accuracy of pre-equilibrium photon estimation might not be improved by its inclusion. Although the freeze-out temperature is set to $T_f = 140$~MeV, the thermal photon contribution after the hydrodynamization before $T = 110$~MeV is taken into account by using the information of hydrodynamic flow calculated after the kinetic freeze-out to partially compensate for the lack of direct photon emission in the hadronic gas stage \cite{Paquet:2015lta}.

The pre-equilibrium and thermal photon emission rates depend on the QCD coupling, the number of colors, and the number of flavors. In this study, they are chosen as $\alpha_s = 0.2$, $N_c = 3$, and $N_f = 3$. The effect of running coupling will be discussed elsewhere.

\subsection{$p_T$ Spectra}

\begin{figure}[tb]
\begin{center}
\includegraphics[width=3.0in]{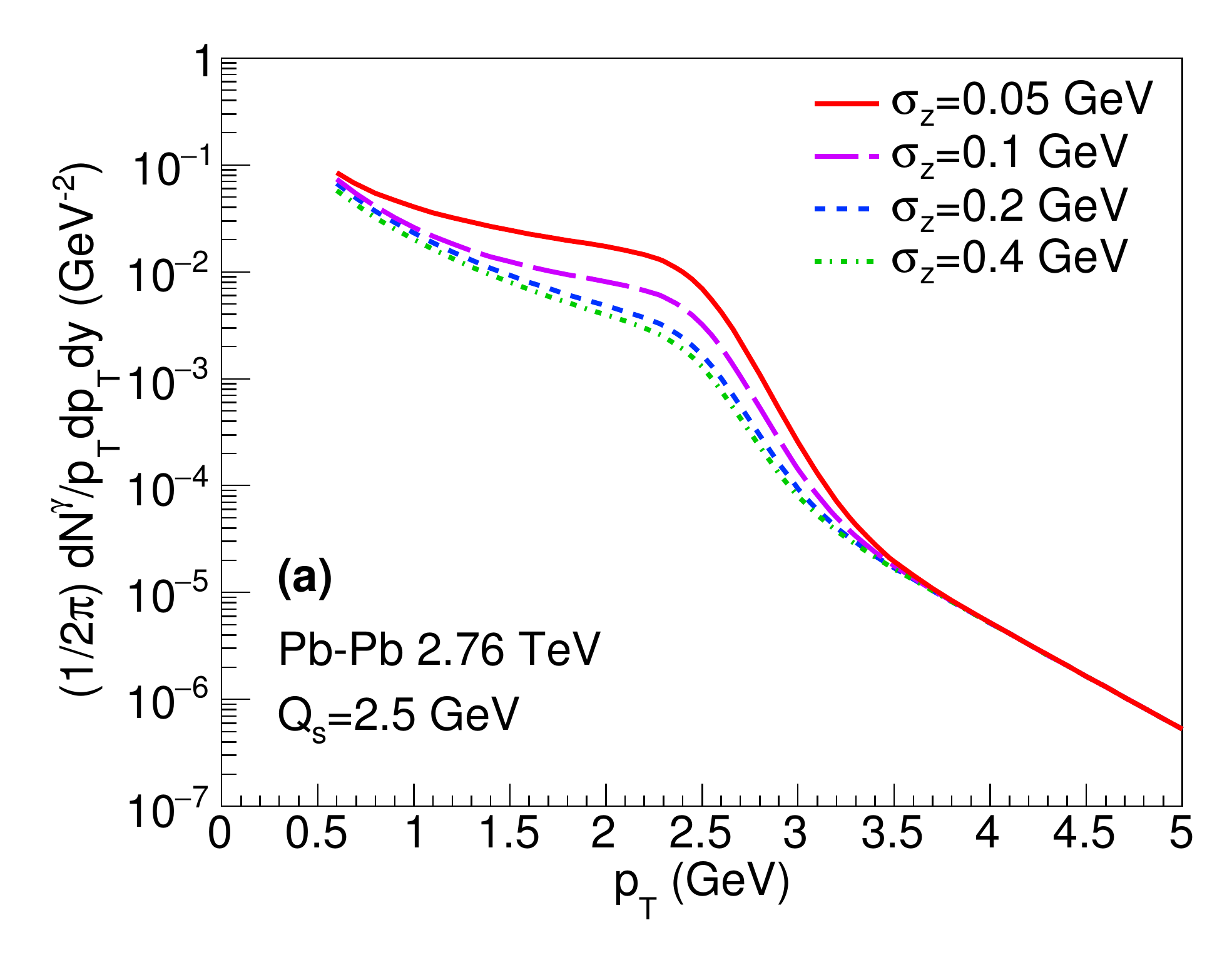}
\includegraphics[width=3.0in]{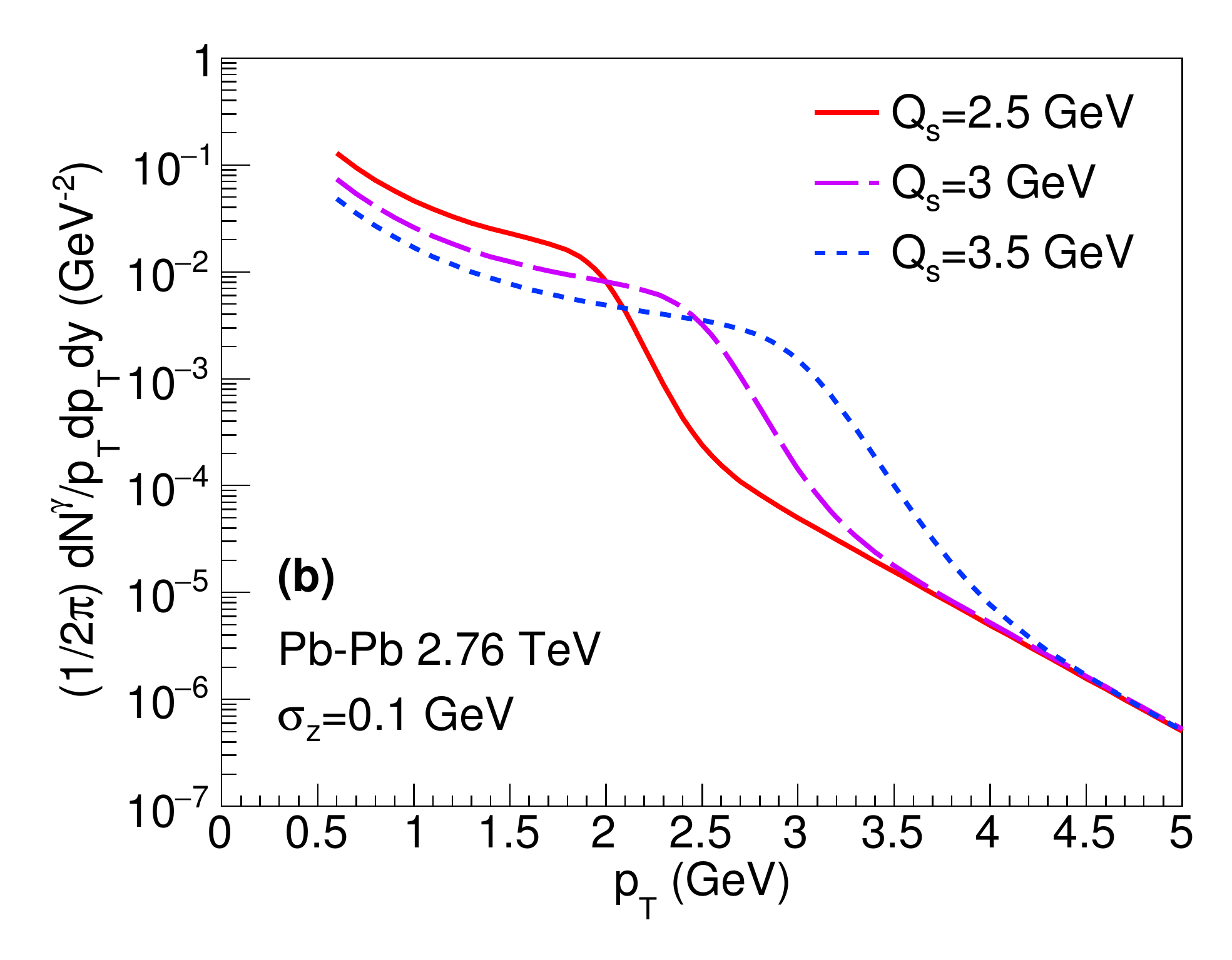}
\end{center}
\caption{(a) $p_T$ spectra of pre-equilibrium photons for $\sqrt{s_{NN}} = 2.76$ TeV Pb-Pb collision at $b = 4.6$ fm with different values of $\sigma_z$ at $Q_s = 2.5$ GeV and (b) those with different values of $Q_s$ at $\sigma_z = 0.1$ GeV.}
\label{fig:1}
\end{figure}

First, particle spectra of direct photons are investigated at midrapidity. $p_T$ spectra of thermal and pre-equilibrium photons are calculated by integrating the respective emission rates over the space-time volume. 

The spectra of pre-equilibrium photons for Pb-Pb collisions are shown in Fig.\ref{fig:1} for different values of the characteristic longitudinal momentum scale $\sigma_z$ and the saturation momentum scale $Q_s$. It should be noted here that $\sigma_z$ and $Q_s$ are the main parameters in the model which need to be explored for minimizing ambiguities. The spectra have a structure around $p_T \sim Q_s$ reflecting the phase-space distributions of partons in the glasma stage. One can see that the emission is larger for smaller values of $\sigma_z$. This is because smaller distribution in the longitudinal momentum space translates into larger distribution in the transverse momentum space for a fixed energy density. Also, the emission at $p_T \sim Q_s$ decreases with increasing $Q_s$ because the energy density is fixed. 

\begin{figure}[tb]
\begin{center}
\includegraphics[width=3.0in]{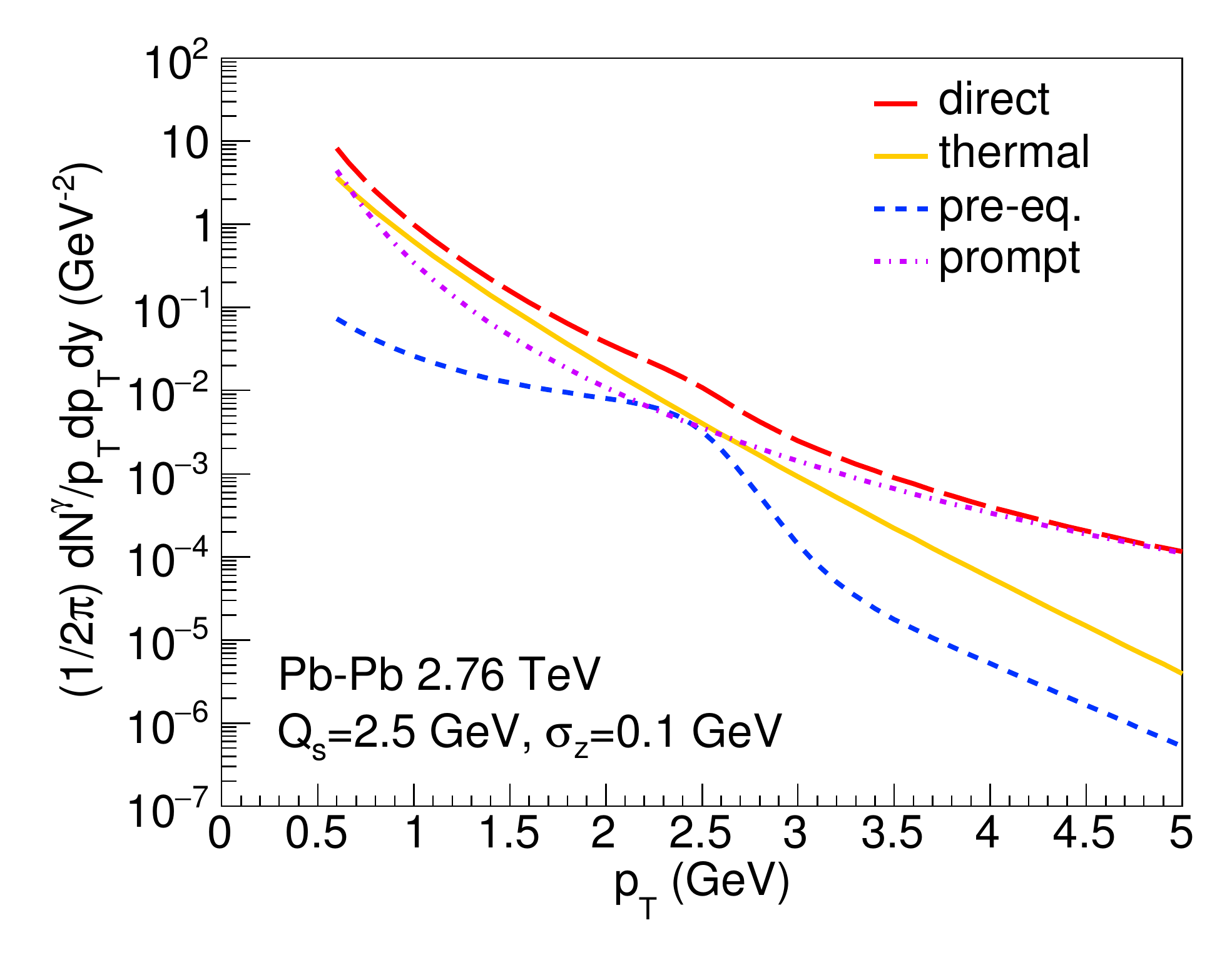}
\end{center}
\caption{$p_T$ spectra of prompt, pre-equilibrium, thermal, and direct photons for $\sqrt{s_{NN}} = 2.76$ TeV Pb-Pb collisions at $b=4.6$ fm. $Q_s = 2.5$ GeV and $\sigma_z = 0.1$ GeV are used.}
\label{fig:2}
\end{figure}

The $p_T$ spectra of prompt photons, pre-equilibrium photons, thermal photons, and direct photons as their sum are shown in Fig.~\ref{fig:2}. Here $Q_s = 2.5$ GeV and $\sigma_z = 0.1$ GeV are chosen for further demonstration. Although the $p_T$-integrated number of pre-equilibrium photons is small owing to the relatively small space-time volume in the system, depending on the anisotropy of the system, its contribution can be comparable to those of thermal and prompt photons at $p_T \sim Q_s$. The pre-equilibrium photons are 47.8\% of the direct photons at $p_T = Q_s = 2.5$ for $\sigma_z = 0.05$ GeV, 29.8\% for $\sigma_z = 0.1$ GeV, 18.0\% for $\sigma_z = 0.2$ GeV, and 14.5\% for $\sigma_z = 0.4$ GeV in the current model. This suggests that pre-equilibrium photons can be non-negligible in relativistic heavy-ion collisions. Prompt photons are relatively more important at higher $p_T$, pre-equilibrium photons at intermediate $p_T \sim Q_s$, and thermal photons at lower $p_T$, reflecting the fact that the typical transverse momentum scale of the system decreases with time evolution.

\begin{figure}[tb]
\begin{center}
\includegraphics[width=3.0in]{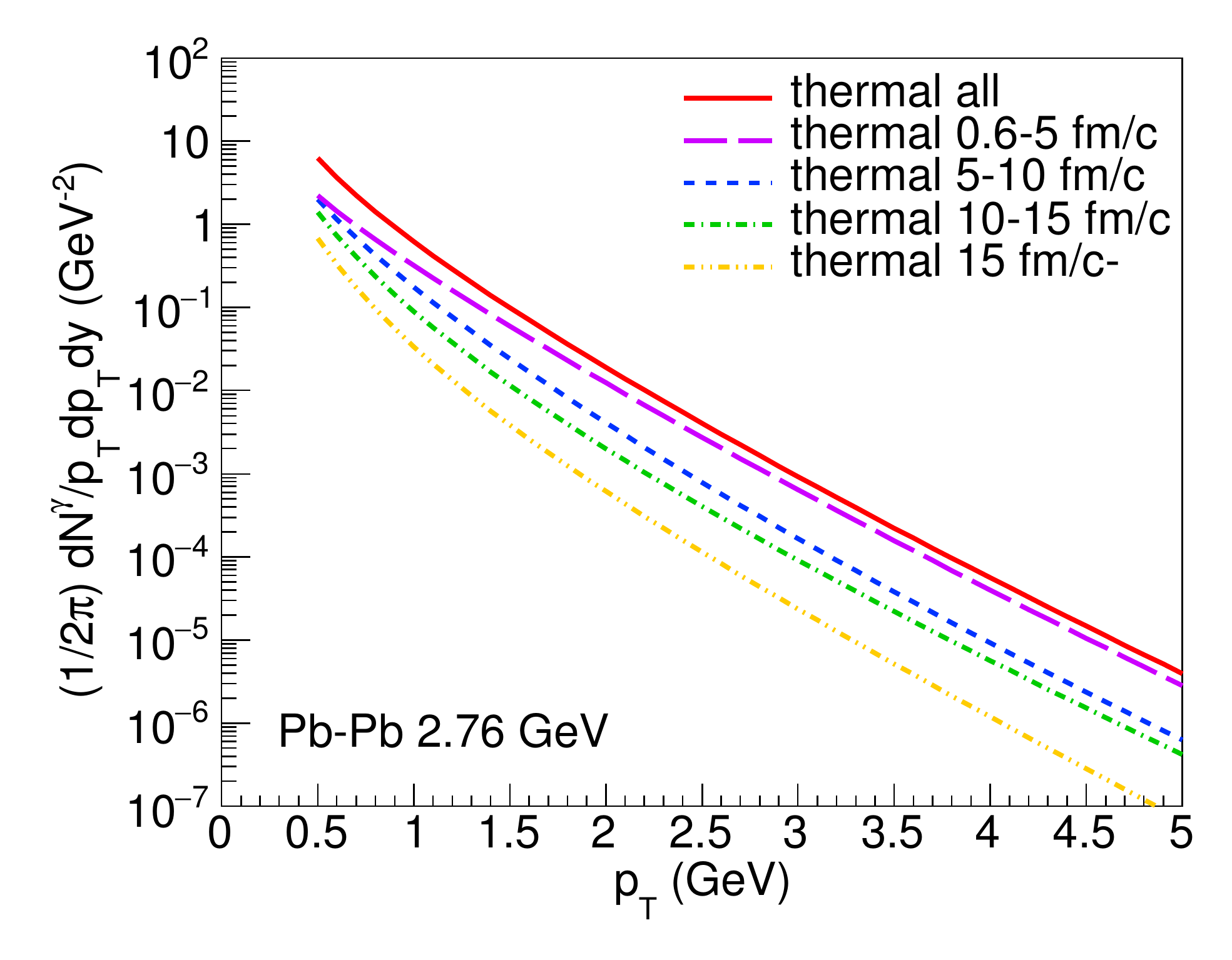}
\end{center}
\caption{$p_T$-spectrum of thermal photons decomposed into the contributions of fixed time intervals for $\sqrt{s_{NN}} = 2.76$ TeV Pb-Pb collisions at $b=4.6$ fm. }
\label{fig:3}
\end{figure}

A similar trend can be found in thermal photon spectra as well. The spectrum is decomposed into the time-dependent contributions in Fig.~\ref{fig:3}.
The slope is harder at earlier times and softer at later times. It can be interpreted as a consequence of the fact that the effective medium temperature is larger at earlier times. The contributions beyond $\tau \sim 15$ fm/$c$ is post-hydrodynamic because the freeze-out hypersurface ends around that time.

It should be noted that the blue-shifting effect of the medium elements which flow toward the observer is partially cancelled by the red-shifting effect of the medium elements on the opposite side that go away from the observer because the medium is optically transparent and roughly axisymmetric. The net effect of the blue-shifting is still positive but the magnitude can be in general smaller than that of the temperature effect.

\subsection{Elliptic flow}

Next, differential elliptic flow of direct photons are investigated to demonstrate the interplay of photon emission from the different stages in the collision. Here the elliptic flow is defined as 
\begin{eqnarray}
v_2^\gamma (p_T,y) &=& \frac{\int d\phi_p \cos[2(\phi_p-\Psi)] \frac{dN^\gamma}{d\phi_p p_T d p_T dy}}{\int d\phi_p  \frac{dN^\gamma}{d\phi_p p_T d p_T dy}} ,
\end{eqnarray}
where $\phi_p$ is the azimuthal momentum angle and $\Psi$ is the event plane angle. $y$ is the rapidity which is zero because midrapidity is considered.

\begin{figure}[tb]
\begin{center}
\includegraphics[width=3.0in]{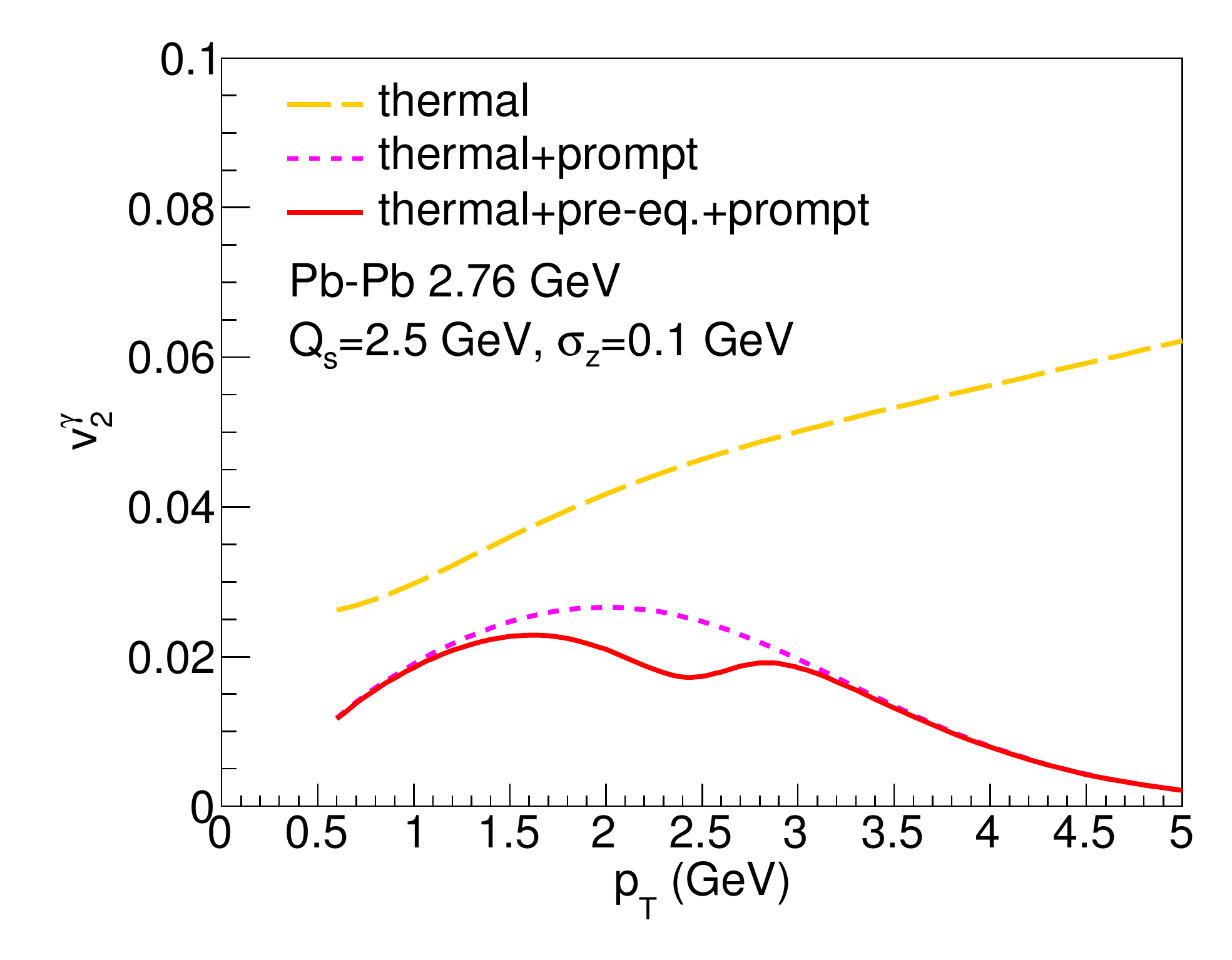}
\end{center}
\caption{$v_2$ of thermal photons, thermal and prompt photons, and direct photons for $\sqrt{s_{NN}} = 2.76$ TeV Pb-Pb collisions at $b=4.6$ fm. $Q_s = 2.5$ GeV and $\sigma_z = 0.1$ GeV are used. }
\label{fig:4}
\end{figure}

The elliptic flow of direct photons is compared with those of thermal photons and of thermal and prompt photons in Fig.~\ref{fig:4}. The prompt and pre-equilibrium photons are assumed to have zero anisotropy in the estimation.
The pre-equilibrium photon contribution tends to reduce direct photon $v_2$ around $p_T \sim Q_s$. 

It is known that most theoretical calculations systematically underestimates direct photon $v_2$ compared with the experimental observation. The inclusion of zero anisotropy pre-equilibrium photons inevitably makes the agreement worse. The results also imply that the flow harmonics of direct photons are sensitive to the details of the physics in the pre-equilibrium stage, and that one has to be careful when assuming zero photon emission from the pre-equilibrium stage.

\section{Discussion and conclusions}
\label{sec4}

The concept of pre-equilibrium photons is introduced to the estimation of direct photons based on the relativistic hydrodynamic model for the first time.
The production of prompt, pre-equilibrium, and thermal photons in relativistic nuclear collisions has been investigated and compared. The thermal photons are estimated by smoothly matching the pQCD and hadronic emission rates. The prompt photons are assumed to be given by the parametrization based on the experimental data of proton-proton collisions. The pre-equilibrium photons are estimated using a model where non-thermal quark and gluon distributions follow a self-similar scaling law, motivated by the turbulent thermalization approach. The original approach is applicable for a weakly-coupled system but here it is assumed to be the dynamics that occurs in the relativistic nuclear collisions by extrapolating and scaling the parameters. The resulting parametric photon emission rate is given as a function of the saturation momentum scale $Q_s$ and the longitudinal momentum scale $\sigma_z$ where the overall normalization is fixed by the experimental data of hadronic particle spectra through the subsequent hydrodynamic model.

The transverse momentum spectra of direct photons have been estimated numerically. It is shown that the effect of pre-equilibrium photons can be non-negligible near $p_T \sim Q_s$, though its magnitude is subject to the arbitrariness in parameter choices. The result implies that if one na\"{i}vely takes the inverse slope parameter of the experimentally-observed spectra to extract the average temperature of the QCD medium, it can be overestimated because it may well be contaminated by the contribution from the stage where the system is not thermalized. Part of the contamination may be cancelled once the full contribution of post-hydrodynamic photons are taken into account.

The elliptic flow of direct photons decreases when the pre-equilibrium photons are included in the estimation assuming that they do not have azimuthal momentum anisotropy. The analyses of the typical momentum range of emission for each source of direct photons, along with those of the time-dependent emission of thermal photons, indicate that early photons are of relevance to the $v_2$ of direct photons at higher momentum where the discrepancy between theoretical estimation and experimental data is larger. This implies that if one simply assumes no photons are produced in the pre-equilibrium stage, the prediction ability of the model could be affected. Though it would be difficult to produce momentum anisotropy in the current model because the non-equilibrium distributions (\ref{eq:preeq_dist1}) are intrinsically isotropic in the transverse directions, it would be interesting to explore how the initial azimuthal momentum anisotropy evolves in the turbulent thermalization picture for studying primordial anisotropy as well as in the early hydrodynamic stage \cite{Kasmaei:2016apv,Kasmaei:2018oag,Kasmaei:2019ofu}. 

The fact that direct photon particle spectra and elliptic flow can have information regarding the pre-equilibrium stage, such as the saturation momentum scale, supports that those observables may be used to experimentally constrain the early time dynamics within the framework of the pre-equilibrium model given that statistics is improved in the experimental data. A similar idea can be found also in Refs.~\cite{Benic:2016yqt,Benic:2016uku,Benic:2018hvb}. Quantitative analyses will be performed in future works.

Future improvements include the introduction of the running coupling to the model. Also, photon emission processes other than pair annihilation and Compton scattering should be considered in the pre-equilibrium stage for more quantitative analyses. The extension of the model to finite chemical potentials \cite{Monnai:2012jc,Monnai:2018rgs,Denicol:2018wdp,Monnai:2019jkc} would be an interesting task, though one has to be careful of the applicability of the color glass condensate picture when using the results for the nuclear collisions at the Beam Energy Scan energies. The quark production process \cite{Blaizot:2014jna,Kasper:2014uaa,Monnai:2014xya} can be non-negligible in the quantum regimes (b) and (c). The numbers of quarks and gluons at the beginning of the pre-equilibrium stage are estimated as $n_q \sim \alpha_s n_g$ aside from the degeneracy factors, reflecting the fact that the system is dominated by gluons in the color glass condensate. The number densities should approach $n_q \sim n_g$ at the end of the stage, which would increase the pre-equilibrium photon emission. Since the chemical equilibration should involve splitting and recombination processes, it can non-trivially modify the phase-space distributions when the thermalization is not fast enough. There also is a possibility that the chemical equilibration process extends into the hydrodynamic stage \cite{Gelis:2004ep,Monnai:2014kqa}. Finally, the direct photon contributions of elastic scattering processes in the post-hydrodynamic stage have to be taken into account for more quantitative understanding of the photon elliptic flow.

\ack
A.M. acknowledges fruitful discussion in Hard Probes 2018 (Aix-les-Bains, France, October 1-5, 2018). 
The work of A.M. was supported by JSPS KAKENHI Grant Number JP19K14722.

\appendix

\section{Pre-Equilibrium Photons in Proton-Proton Collisions}
\label{sec:A}

\begin{figure}[tb]
\begin{center}
\includegraphics[width=3.0in]{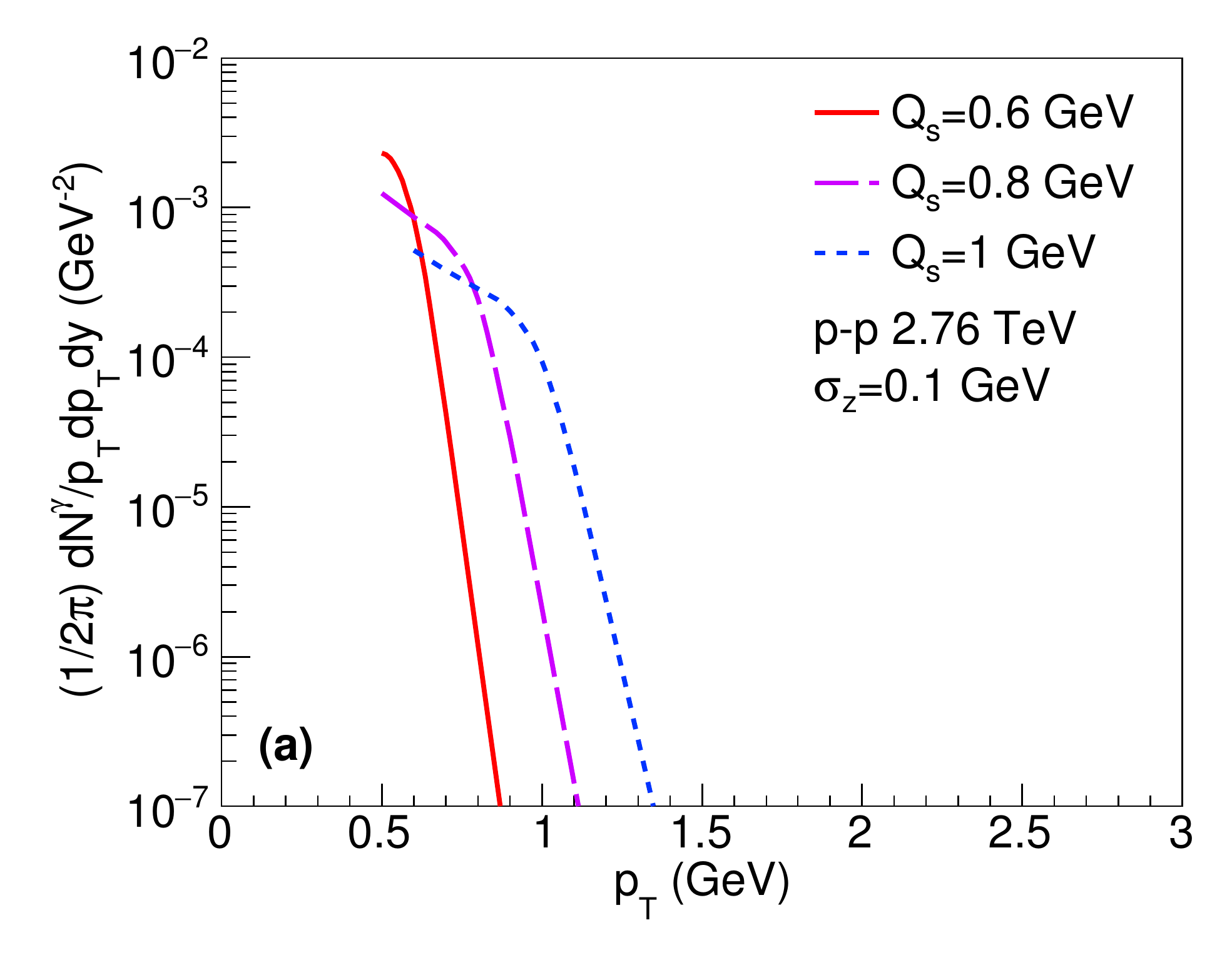}
\includegraphics[width=3.0in]{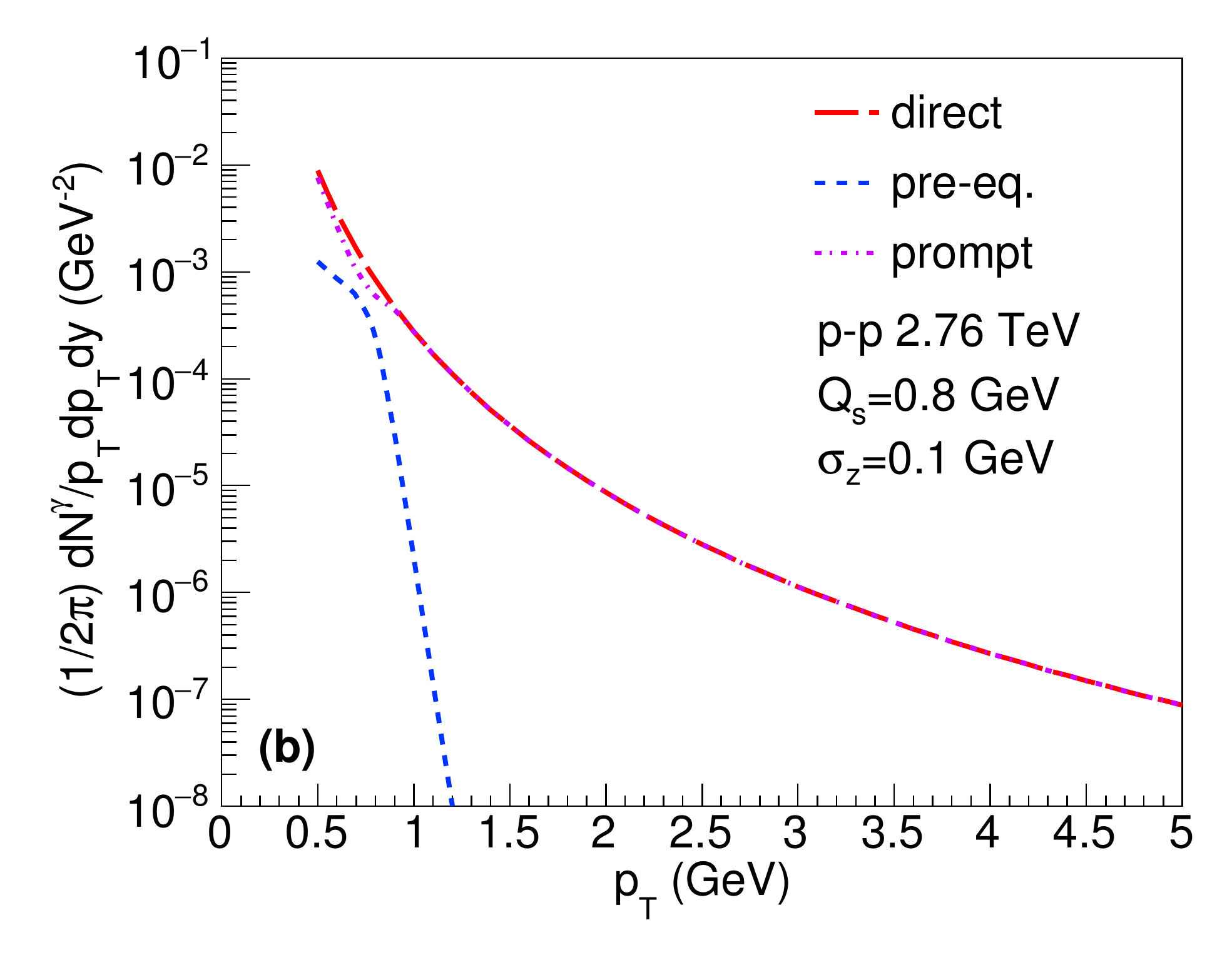}
\end{center}
\caption{(a) $p_T$ spectra of pre-equilibrium photons for $\sqrt{s_{NN}} = 2.76$ TeV proton-proton collisions at $\sigma_z = 0.1$ GeV. (b) Those of prompt, pre-equilibrium, and direct photons at $Q_s = 0.8$ GeV. }
\label{fig:5}
\end{figure}

\begin{figure}[tb]
\begin{center}
\includegraphics[width=3.0in]{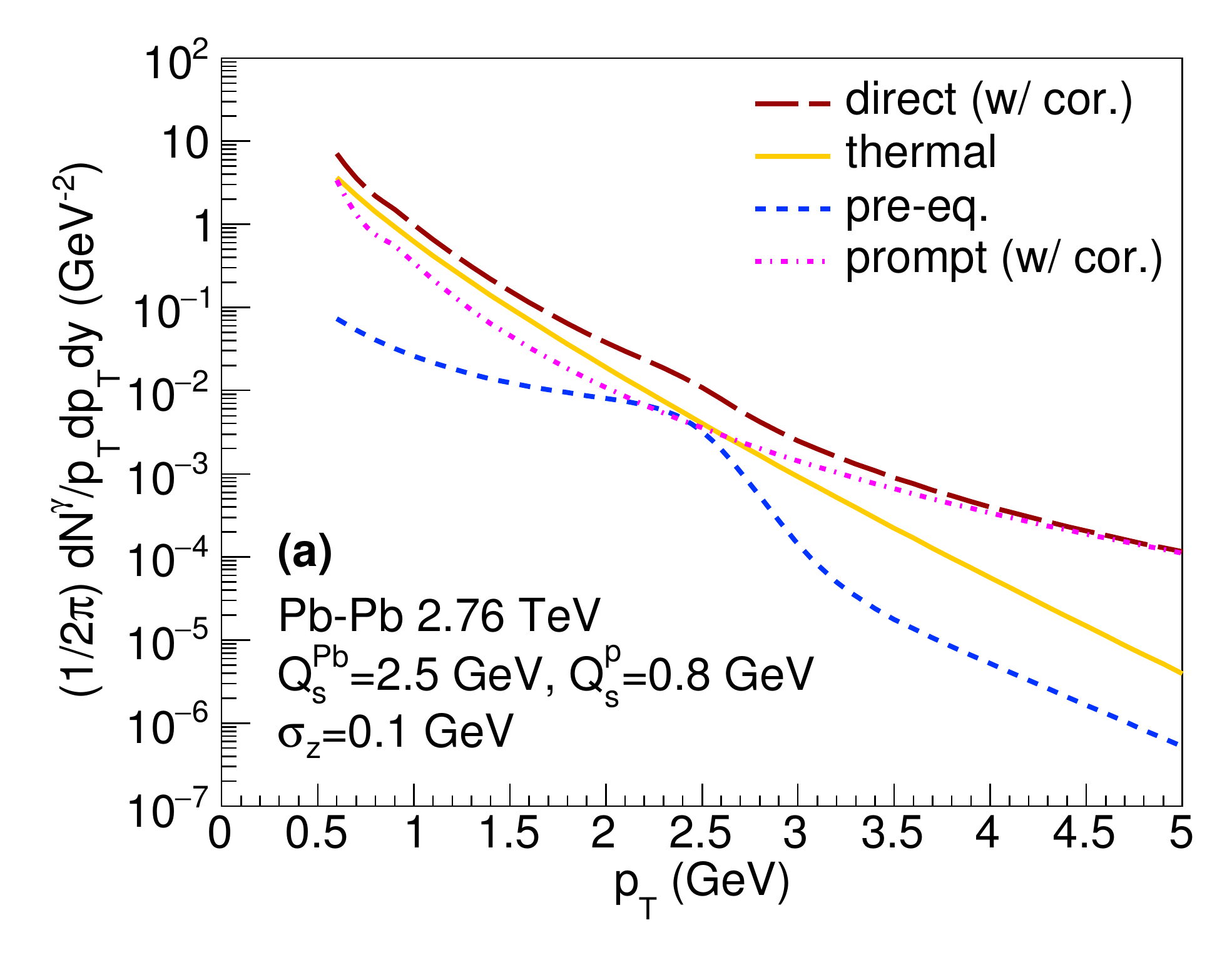}
\includegraphics[width=3.0in]{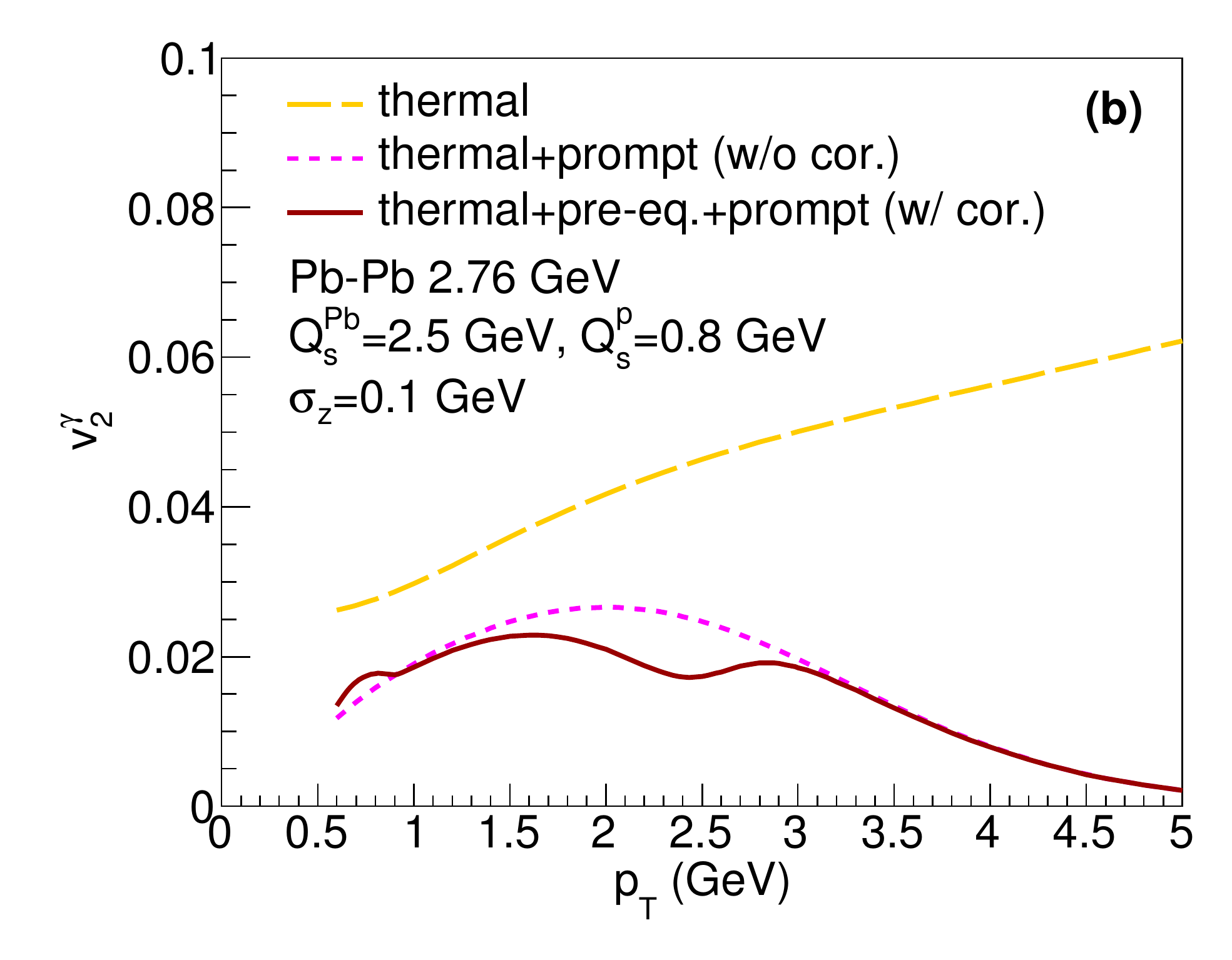}
\end{center}
\caption{(a) $p_T$ spectra of glasma-corrected prompt photons with those of pre-equilibrium, thermal, and direct photons for $\sqrt{s_{NN}} = 2.76$ TeV Pb-Pb collisions at $b=4.6$ fm. $Q_s^\mathrm{Pb} = 2.5$ GeV, $Q_s^\mathrm{p} = 0.8$ GeV, and $\sigma_z = 0.1$ GeV are used. (b) $v_2$ of thermal photons compared with those with glasma-uncorrected prompt photon contrition and with glasma-corrected prompt photon and pre-equilibrium photon contributions.}
\label{fig:6}
\end{figure}

Recent discoveries of possible primordial collectivity in small systems \cite{PHENIX:2018lia} implies the existence of non-prompt photon contributions in such collisions. Apart from prompt photons, pre-equilibrium photons can be produced in proton-proton collisions at top LHC energies because the saturation momentum scale is expected to approach $Q_s \sim 1$~GeV \cite{GolecBiernat:1998js} and a glasma-like structure may well be produced. Recent analyses of the small systems even suggest that they may follow hydrodynamic description \cite{Weller:2017tsr}, which hints at the production of primordial thermal photons \cite{Shen:2016zpp}. 
One can obtain the pre-equilibrium parton distributions following the same procedure as the one for Pb-Pb collisions.

As the existence of a thermalized medium in proton-proton collisions is still under debate \cite{Mace:2018vwq}, prompt and glasma photons are considered as the components of direct photons. Note that the normalization of the pre-equilibrium phase space distribution is still constrained so that if the hydrodynamic model were used it could reproduce the hadronic particle spectra up to $p_T \sim 1$~GeV with $T_f = 170$ MeV. Using the expression in Sec.~\ref{sec:preeq}, the direct photon spectra is expressed as
\begin{eqnarray}
E \frac{dN^\gamma_\mathrm{dir}}{d^3p} &=& E \frac{dN^\gamma_\mathrm{pro}}{d^3p} + \int dx^4 E \frac{dR^\gamma_a}{d^3p},
\end{eqnarray}
where $R^\gamma_a$ is the photon emission rate in proton-proton collisions during the stage (a). Since local equilibration of the system is not assumed in the estimation, the contribution from the softer stages (b) and (c) are not taken into account for a conservative estimate. 

The direct photon estimation based on the experimental measurements should be considered as the sum of prompt and pre-equilibrium photon contributions. On the other hand, a na\"{i}ve pQCD estimation is related only to the former. The two estimations are shown to agree down to around $p_T \sim 1$~GeV at the RHIC energies \cite{Paquet:2015lta}. However, it is not clear if the agreement continues to hold below 1~GeV because the pQCD calculations would break down at lower $p_T$ and the corresponding data is also not yet available in high-energy proton-proton collisions. The existence of pre-equilibrium photons suggest that the two estimations can eventually disagree below $p_T \sim Q_s$.

The glasma photon spectra for different values of $Q_s$ are shown in Fig.~\ref{fig:5} (a). As is the case for the heavy-ion collisions, the spectra have a characteristic structure near $p_T \sim Q_s$. Since the contribution from the classical stage is considered, the primordial thermal tail is not observed.
$p_T$ spectra of prompt and pre-equilibrium photons are shown in Fig.~\ref{fig:5} (b) assuming  that of direct photons is given by the formula (\ref{eq:ppscaled}). One has to be careful that this is a conjecture because the parametrization is based on the extrapolation of the experimental data available down to around 2~GeV. Here $Q_s = 0.8$ GeV is used for demonstration. The prompt photon spectra is then determined as the difference of the direct and pre-equilibrium photon spectra.

The modification of prompt photon estimation can have an effect on the analyses of heavy-ion collisions. Figure \ref{fig:6} (a) shows the particle spectra of prompt photons for Pb-Pb collisions at $b=4.6$~fm with the correction of pre-equilibrium photons in proton-proton collisions. Comparing with the previous results in Fig.~\ref{fig:2}, prompt photons and consequently direct photons are slightly reduced near $p_T \sim Q_s^\mathrm{p}$ for the current parameter set. It makes the inverse slope slightly flatter. The differential elliptic flow is shown in Fig.~\ref{fig:6} (b). When the pre-equilibrium photons are considered both in Pb-Pb and in proton-proton collisions, the net direct photon momentum anisotropy is reduced near $p_T \sim Q_s^\mathrm{Pb}$ and enhanced near $p_T \sim Q_s^\mathrm{p}$. 
The structures may be less apparent in the experimental data because the pre-equilibrium and hydrodynamic stages may not be clearly separated in actual relativistic nuclear collisions. Those results motivates one to further investigate low-momentum direct photon spectra in proton-proton collisions once the data become available.

\section*{References}

\bibliography{dir_ph}

\end{document}